\documentclass[page-classic]{epl2}
\usepackage{graphicx}

\title{Rod-like Polyelectrolyte Brushes with Mono- and Multivalent Counterions}

\author{H. Fazli$^1$, R. Golestanian$^1$, P.L. Hansen$^2$, \and M.R. Kolahchi$^1$}

\institute{$^1$ Institute for Advanced Studies in Basic Sciences,
Zanjan 45195-1159, Iran\\
$^2$ MEMPHYS-Center for Biomembrane Physics, Physics
Department, University of Southern Denmark, DK-5230 Odense M,
Denmark}




\abstract{A model of rod-like polyelectrolyte brushes in the presence of
monovalent and multivalent counterions but with no added-salt is
studied using Monte Carlo simulation. The average height of the
brush, the histogram of rod conformations, and the counterion
density profile are obtained for different values of the grafting
density of the charge-neutral wall. For a domain of grafting
densities, the brush height is found to be relatively insensitive
to the density due to a competition between counterion
condensation and inter-rod repulsion. In this regime, multivalent
counterions collapse the brush in the form of linked clusters.
Nematic order emerges at high grafting densities, resulting is an
abrupt increase of the brush height.}
\begin{document}
 \maketitle

Electrostatic correlation effects of highly charged macro-ions in
aqueous solutions have recently attracted much attention
\cite{Moreira}. Novel phenomena such as charge inversion and
like-charge attraction, which cannot be understood within the
Poisson-Boltzmann theory, appear to play a key role in biological
processes such as DNA packaging \cite{DNA} and cell scaffolding
dynamics \cite{cyto}. Such correlations, and their role in
determining the overall structural properties of the system, are
most pronounced in high density polyelectrolyte solutions where
the charge ``patterns'' that form due to the correlations strongly
interact with each other. If the polyelectrolyte is sufficiently
stiff in its backbone structure [such as DNA and filamentous actin
(F-actin)], a geometric constraint is imposed on the charges in
each polyelectrolyte segment to adopt a more or less linear
configuration, which makes a solution of such rod-like
polyelectrolytes a patterned matrix for the counterions (that is,
the ions of opposite charge that are present to neutralize the
overall solution). The interplay between these correlations and
patterns can lead to novel structural and dynamical properties
\cite{gerard,Gelbart,Bruinsma,Fazli}, the study of which could be
useful in understanding the biological phenomena mentioned above
as well as designing functional biomimetic materials.

A common realization of such high density assemblies occurs in
polyelectrolyte brushes \cite{PEbrush}, where charged polymers are
end-grafted to surfaces. While polyelectrolyte brushes have been
mostly studied because of their role in stabilization of colloidal
suspensions, understanding their structural properties could help
us in a variety of problems in biophysics and -technology. A case
in point is filamentous microtubules, {\em i.e.} rod-like
structures build from tubulin monomers with charged C-terminal
amino-acid tails extending from the filament core \cite{Sataric}.
The C-terminal tails seem to play key roles in microtubule
function, e.g. in connection with binding of
microtubule-associated proteins and processivity of the molecular
motors that move along microtubules \cite{MT}. Presumably they
form disordered brushes making it difficult to study them by
systematic scattering and osmotic stress experiments
\cite{Safinya}. Design of more effective DNA chips \cite{DNAchips}
may also benefit from a better understanding of rod-like
polyelectrolyte brushes. Interestingly, STM studies suggest that
densely grafted arrays of short single-stranded DNA molecules on
gold substrates self-organize into disordered structures, and it
takes an applied electric field normal to the grafting surface to
align these brushes \cite{Mikala}.

\begin{figure}
\onefigure[width=14cm]{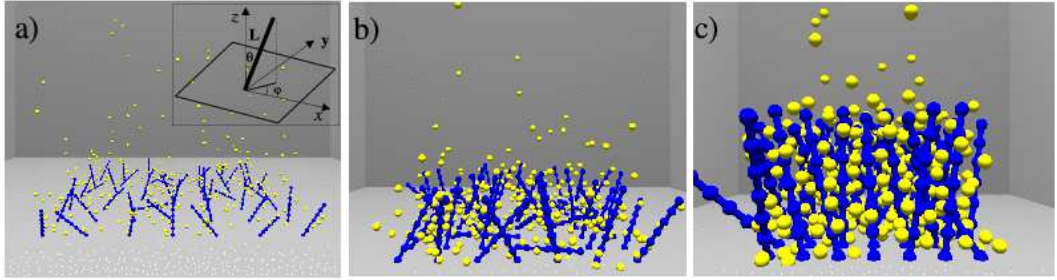} \caption{Snapshots of the system
with monovalent counterions: a) $a=1.2 L$, b) $a=0.55 L$ and c)
$a=0.25 L$. Inset: the schematic configuration of a
polyelectrolyte rod of length $L$ and two rotational degrees of
freedom $\theta$ and $\phi$.}\label{fig:snapshot1}
\end{figure}

\begin{figure}
\onefigure[width=10cm]{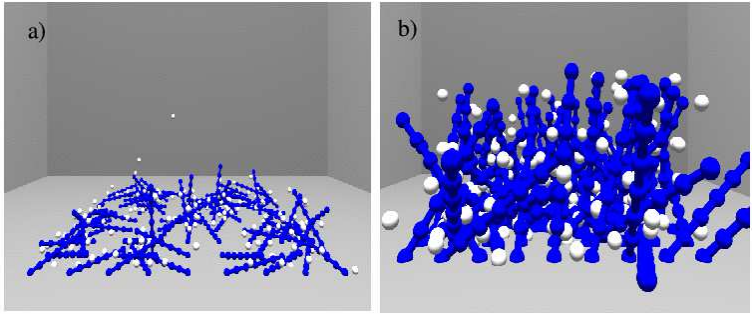} \caption{Snapshots of the system
with trivalent counterions at a) $a=0.65 L$ and  b) $a=0.25 L$.}
\label{fig:snapshot2}
\end{figure}

Most theoretical studies of polyelectrolyte brushes have focused
on flexible chains, i.e. charged polymers that are longer than
their persistence lengths. If the polyelectrolytes are
sufficiently short or highly charged, they would behave
effectively as charged rods \cite{Tirrel} that are constrained to
be very close to each other because of the brush structure; a
feature that can bring about frustration \cite{Fazli}. Here, we
study the structural properties of rodlike polyelectrolyte brushes
in the presence of monovalent and trivalent counterions using
Monte Carlo (MC) simulation techniques. We calculate the average
thickness of the brush, the density profile of the counterions,
and the statistics of the conformation of the rods, as functions
of the surface coverage density. We find three distinct regimes
for the collective behavior of the rods: (i) a weakly interacting
regime at low densities where the counterions are dispersed in the
solution (see Fig. \ref{fig:snapshot1}a), (ii) an intermediate
regime where the counterions are condensed in the brush (see Fig.
\ref{fig:snapshot1}b), and (iii) a novel nematic regime at high
densities (see Fig. \ref{fig:snapshot1}c). The counterion density
profile is found to be depleted near the wall in the intermediate
regime, while the statistics of the rod conformations show that a
small fraction of the rods choose to lie on the surface as a
result of electrostatic frustration. We also find that trivalent
counterions collapse the brush in the intermediate regime by
forming links between clusters of neighboring rods (see Fig.
\ref{fig:snapshot2}a). Interestingly, the height of the brush in
the intermediate regime seems to depends very weakly on the areal
density of the rods, both for monovalent and trivalent
counterions.

We model each polyelectrolyte as a rigid rod of length $L$ with
$N_m$ charged hard sphere monomers of diameter $\sigma$ and charge
$q_m=-e$ distributed along it in a uniform periodic array. $N_r$
polyelectrolytes are end grafted on a square lattice with a
lattice constant $a$ on a surface, so that the areal grafting
density is $\rho_a=1/a^2$. Each polyelectrolyte has two rotational
degrees of freedom $\theta$ and $\phi$ (Fig.
\ref{fig:snapshot1}a). The counterions are modeled as hard-core
spheres of diameter $\sigma$ and they carry a charge $q_c=Z e$ in
which $Z$ is the valence of the counterions. The system is
electrostatically neutralized: $N_c\times Z=N_r\times N_m$, where
$N_c$ is the number of counterions.

We use Lekner summation method \cite{Lekner} to account for
long-range Coulomb interactions by applying lateral periodic
boundary conditions in $x$ and $y$ directions (see Fig.
\ref{fig:snapshot1}a), and the simulation box is finite in $z$
direction, normal to the brush surface. Elementary moves for
counterions, are chosen at random in the simulation box around it.
For the polyelectrolyte rods the non-grafted end of each rod moves
randomly on the surface of a hemisphere of radius $L$. For highly
charged rods it is known that algorithms involving independent
moves for the polyelectrolytes become inefficient since many
counterions are linked to polyelectrolytes, and moving a rod away
from these counterions would cost a lot of energy that effectively
prohibits any Monte Carlo move \cite{Belloni}. To solve this
problem, we follow Ref. \cite{Belloni} and look for all
counterions at distances less than $d$ of each rod and rotate them
all together with the polyelectrolyte. Considering the acceptance
rate of the MC moves, the value of $d$ has been optimized to $d=2
\sigma$. We equilibrate the system by starting from different
initial configurations and reaching the same stationary value of
the Coulomb energy. In simulations with monovalent counterions,
the equilibrium was reached after $2 \times 10^4$ MC steps per
particle for the smallest value of the lattice constant. For the
case of trivalent counterions the equilibration time is $3 \times
10^4$ MC steps per particle. The simulation box height in $z$
direction, $H$, has been chosen in such a way that the results are
independent of its value; $H=12 L$ for monovalent and $H=8 L$ for
trivalent counterions.

To quantify the strength of the electrostatic energy relative to
the thermal energy, we can define the Bjerrum length $l_{\rm
B}={{e^2}\over{\epsilon k_{\rm B}T}}$, where $\epsilon$ is the
dielectric constant of the solvent. In our simulations, we chose
to fix the Bjerrum length---which in room temperature is about 0.7
nm for water---to a fraction of the length of the rod, namely,
$l_{\rm B}=L/7$. This means that the length of rods is chosen to
be about $L \simeq 5$ nm at room temperature. We also fixed the
diameter of the rods to $\sigma=0.1 L$. In the case of monovalent
counterions, the simulation parameters were: $N_m=7$, $N_r=7
\times 7=49$, and $N_c=343$. Thermal averages were obtained over
$6 \times 10^4$ MC steps, after equilibrating the system. In
simulations with trivalent counterions ($Z=3$), we chose: $N_m=7$,
$N_r=9 \times 9=81$, and $N_c=189$, and thermal averages are
calculated over $9 \times 10^4$ MC steps after equilibration. Note
that the Manning parameter $\xi=\frac{Z N_m l_{\rm B}}{L}$
\cite{Manning} was set to $\xi_1=1$ for monovalent, and $\xi_3=3$
for trivalent counterions.

\begin{figure}
\twofigures[width=7cm]{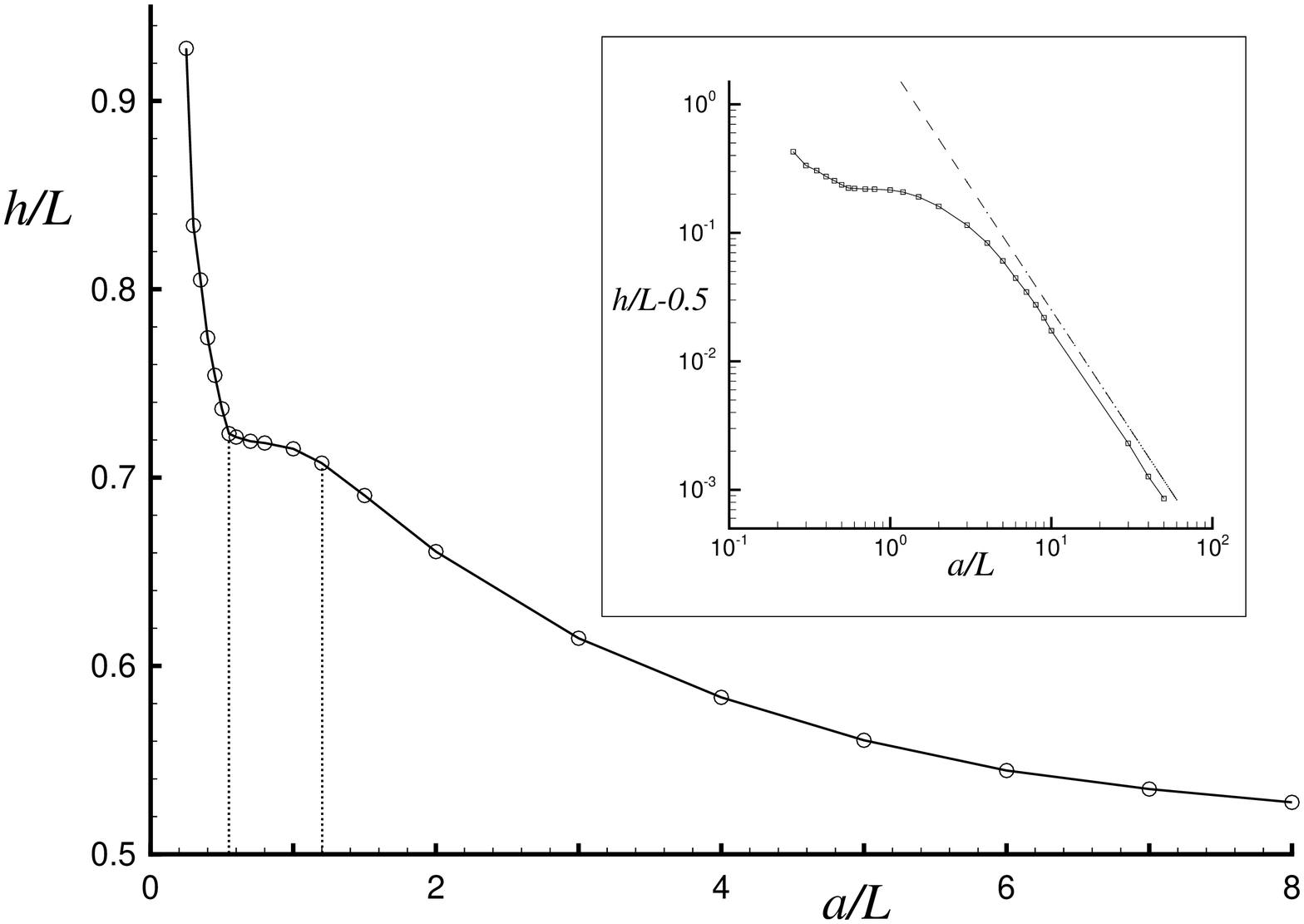}{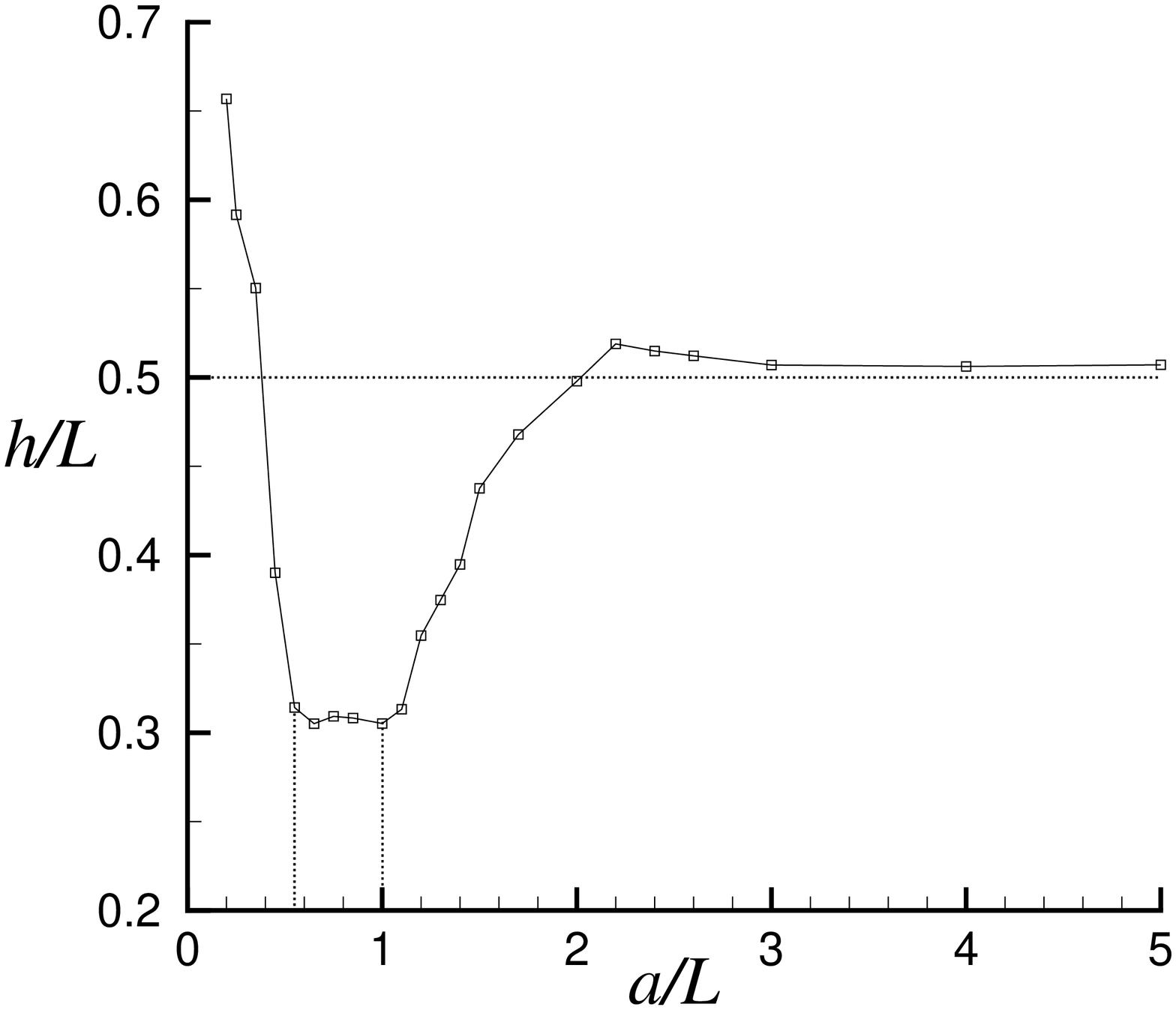} \caption{The brush
height as a function of the lattice constant $a$, in the presence
of monovalent counterions. Inset: the log-log plot of $h/L-0.5$ as
a function of $a$ for $Z=1$. At large lattice constants the brush
height scales as $a^{-2}$ (the dashed line has the slope of
$-2$).} \label{fig: brush_height-Z1} \caption{The brush height
versus $a$ in the presence of trivalent counterions.} \label{fig:
brush_height-Z3}
\end{figure}

We vary the lattice constant $a$, keeping the other parameters
fixed, and calculate the brush height as
\begin{math}
h=\frac{L}{N_r}\sum_{i=1}^{N_r}\left\langle
\cos\theta_{i}\right\rangle,
\end{math}
where $\left\langle \cdots \right\rangle $ denotes ensemble
average. For each value of $a$, we start the system from a random
initial configuration and calculate thermal averages after
equilibration of the system. Figure 3 shows the brush height as a
function of the lattice constant $a$ for $Z=1$. When the lattice
constant is very large, the effective attraction of the rods to
the counterions becomes very weak and the counterions evaporate
away from them, while the long distance between the rods makes
their own electrostatic interaction negligible. In this regime,
$\cos\theta$ has a uniform distribution with the thermal average
of $\frac{1}{2}$, and the brush height is $h=L/2$. As the spacing
$a$ is reduced, the rods start to interact with each other
electrostatically and the brush is swollen, as can be seen in Fig.
\ref{fig: brush_height-Z1}. Note that in this regime the majority
of the counterions are still away from the surface (see the
snapshot of the system in Fig. \ref{fig:snapshot1}a). To justify
this observation, one can roughly consider the $z=0$ plane as a
charged surface with the unform charge density
$\sigma_s=\frac{N_m}{a^2}$, for which it is known from
Poisson-Boltzmann theory that the counterions are confined within
a distance of $\lambda=\frac{1}{2\pi Z l_{\rm B}
\sigma_s}=\frac{a^2}{2\pi Z l_{\rm B} N_m}$, called the
Gouy-Chapman length. We can readily see that $\lambda \gg L$, so
long as $a \gg L$ and $\lambda \gg a$ that is a requirement for
the smeared surface charge density approximation to be justified.

The swelling of the brush in this regime seems to obey a scaling
law of the form $\frac{h}{L}-\frac{1}{2} \sim \frac{1}{a^2}$, as
demonstrated in the inset of Fig. \ref{fig: brush_height-Z1}. This
scaling behavior can be understood by assuming that individual
rods are fluctuating as dipoles in the presence of a uniform
electric field as obtained from the smeared charge density
$\sigma_s$, namely, the sum of monopole-dipole interactions which
are the dominant terms in the multipole expansion valid for large
$a$. The scaling behavior is reminiscent of that of flexible
polyelectrolyte brushes in the Pincus regime \cite{PEbrush}.

Upon decreasing the lattice constant, the effective surface charge
density of the brush is increased so that the counterions are
attracted to the brush, as the snapshot of the system in Fig.
\ref{fig:snapshot1}b shows. This trend continues until a new
regime sets in where all the counterions are confined within the
brush, which is analogous to the osmotic regime in flexible
polyelectrolyte brushes \cite{PEbrush}.  One could roughly
estimate the onset of this crossover by setting the thickness of
the counterion layer ($\sim \lambda$) equal to the height of the
brush ($\sim L$), which yields $a_{\times}\sim \xi^{1/2} L$. In
this regime, the counterions are involved in heavily screening the
interaction between the rods, and thus one expects that the
swelling slows down, as can be seen in the plateau-like feature in
Fig. \ref{fig: brush_height-Z1} for the intermediate range of
$0.55 L < a < 1.2 L$. At higher densities, the rods are completely
neutralized by the counterions, and the remaining excluded volume
interaction among them promotes a novel nematic ordering that sets
in sharply at $a=0.55 L$ (see Fig. \ref{fig: brush_height-Z1}),
similar to the Onsager nematic transition in 3D solutions of hard
rods \cite{Nematic}. A typical snapshot of the system in the
nematic phase is shown in Fig. \ref{fig:snapshot1}c.

The picture becomes more complicated with the introduction of
multivalent counterions. Figure \ref{fig: brush_height-Z3} shows
the brush height as a function of the lattice constant in the
presence of trivalent counterions. For infinitely large $a$ the
the brush height is still $h=L/2$, and it does swell initially as
$a$ is decreased similar to the monovalent case. However, once a
sufficient fraction of the trivalent counterions are attracted to
the rods and fluctuate in the vicinity of them, they give rise to
a fluctuation--induced attraction \cite{FII}. This attractive
interaction competes with the repulsion between the rods and the
result of this competition turns back the swelling of the brush
around $a=2.2 L$ (see Fig. \ref{fig: brush_height-Z3}). This trend
becomes more drastic as of $a=2.0 L$, where the brush height is
$h=0.5 L$ again, because the trivalent counterions can now link
the polyelectrolyte rods to each other (see Fig.
\ref{fig:snapshot2}a) and collapse the brush \cite{Santan}.
Similar to the monovalent case, an interesting plateau regime
appears for $0.55 L < a < 1.0 L$, where the enhancement in the
repulsion between the rods as $a$ decreases is exactly balanced by
the increased possibility of forming links between neighboring
rods. This leads to the clusters growing in size more and more
until at $a=0.55 L$ the collection of now fully neutralized hard
rods undergo a nematic transition; the brush height suddenly
increases (see Fig. \ref{fig: brush_height-Z3}) and the rods
become predominantly parallel, as the snapshot in Fig.
\ref{fig:snapshot2}b shows. Note that the onset of the nematic
transition coincides in the two cases of mono- and trivalent
counterions, in line with the suggestion that the transition is
entirely driven by the excluded volume interaction.

\begin{figure}
\onefigure[width=13cm]{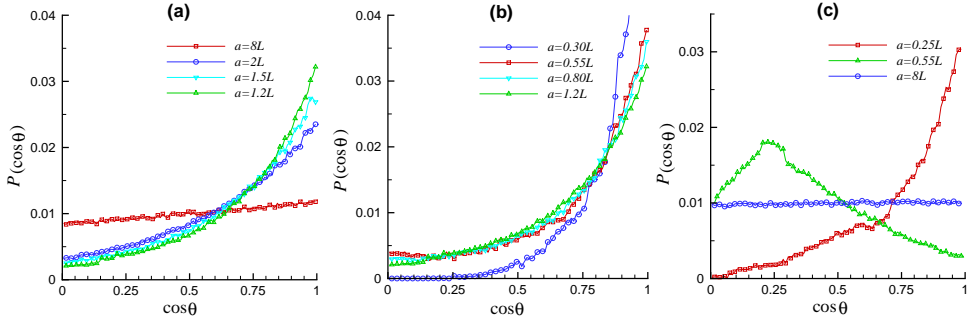} \caption{The histogram of
$\cos\theta$ for a) $Z=1$ and $a\geq 1.2 L$, b) $Z=1$ and $a\leq
1.2 L$, and c) $Z=3$.} \label{fig:histogram}
\end{figure}

In addition to the average thickness of the brush, it is
instructive to monitor the histogram of the rod conformations
$P(\cos\theta)$ in the different regimes. In Fig.
\ref{fig:histogram}a, $P(\cos\theta)$ is shown for four different
lattice constants in the $a \geq 1.2 L$ domain, for monovalent
case. We find that in this regime by decreasing $a$, the
$\cos\theta=0$ part of the histogram decreases and the
$\cos\theta=1$ part increases, which corresponds to a ``normal''
swelling of the brush. (At sufficiently large lattice constants
the histogram shows a uniform distribution of $\cos\theta$, as
expected.) As the lattice spacing is further reduced to $a\leq 1.2
L$, the situation becomes more subtle, as Fig.
\ref{fig:histogram}b shows. For $0.55 L < a < 1.2 L$ the trend of
change in the $\cos\theta=0$ part of the histogram is different
from Fig. \ref{fig:histogram}a: decreasing the lattice constant
$a$ increases both the $\cos\theta=0$ and $\cos\theta=1$ parts of
the histogram, while the intermediate region of the histogram
decreases to maintain the normalization of the distribution. In
this regime, the system presumably tries to lower the
electrostatic repulsion between the polyelectrolytes by increasing
angles between the neighboring rods and consequently some of the
rods lie on the surface, as can be seen in the snapshot for
$a=0.55L$ in Fig. \ref{fig:snapshot1}b. For $a< 0.55 L$, a drastic
change of behavior is observed and the $\cos\theta=0$ part of the
histogram vanishes, which signals the nematic ordering of the
rods. The histogram is also shown for the case of trivalent
counterions in Fig. \ref{fig:histogram}c. While the histogram is
similar to the monovalent case in the large-$a$ regime and in the
nematic phase, for intermediate values of $a$ it develops a
maximum at some angle that is characteristic of the clusters that
form when the multivalent counterions link neighboring rods (see
Fig. \ref{fig:snapshot2}a).

\begin{figure}
\onefigure[width=13cm]{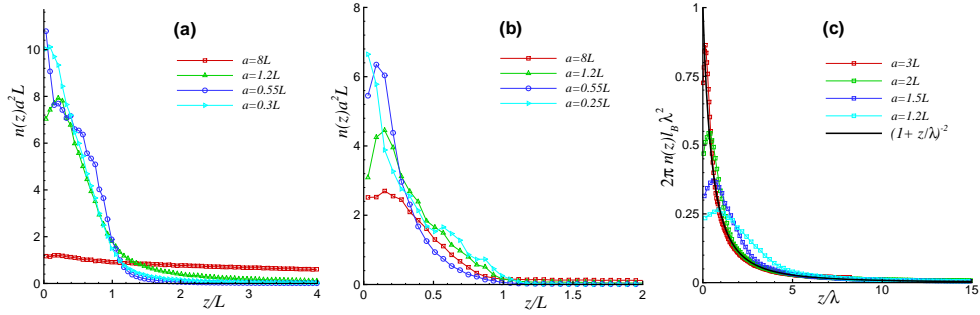} \caption{The number density of
counterions in units of $a^{-2}L^{-1}$ as a function of $z/L$ for
different values of $a$ for a) $Z=1$ and b) $Z=3$. c) Rescaled
number density of monovalent counterions as functions of
$z/\lambda$. } \label{fig:density_profile}
\end{figure}

The histogram of $\cos\theta$ can be used to determine the average
density profile of the rods. One can define the fluctuating
density ${\rho}_{\rm Rod}({\bf r})=\frac{N_m}{L} \sum_{i} \int_0^L
ds \;\delta^3({\bf r}-{\hat {\bf t}}_{i} s)$ where ${\hat {\bf
t}}_{i}$ denotes the unit director of the $i$-th rod, and the rods
are assumed to have a smeared charge distribution for simplicity
of the presentation. We define the average density profile as
$n_{\rm Rod}(z)=\frac{1}{N_r a^2} \int dx dy \left\langle
{\rho}_{\rm Rod}({\bf r})\right\rangle$, where the averaging over
the direction of the rods can be performed using the distribution
$P(\cos\theta)$. After some straightforward manipulation, one can
write the density profile as
\begin{math}
n_{\rm Rod}(z)=\frac{N_m}{L a^2} \int_{z/L}^1 \frac{d t}{t} \;
P(t).
\end{math}
This expression shows that the density profile of the rods will
have a singular behavior near the surface if the $\cos \theta=0$
part of the distribution is nonzero, {\em i.e.} a non-vanishing
fraction of the rods lie on surface.

We have also obtained the counterion density profile as a function
of $z$ for each value of $a$, as shown in Fig.
\ref{fig:density_profile}a for $Z=1$, and in Fig.
\ref{fig:density_profile}b for $Z=3$. For all values of the
spacing, the general expectation that the counterions are confined
within a distance of the order the Gouy-Chapman length holds true.
This is demonstrated in Fig. \ref{fig:density_profile}c where the
profiles are compared with the exact solution of the
Poisson-Boltzmann equation for a flat uniformly charged surface
\begin{math}
n(z)=\frac{1}{2\pi Z^2 l_B
\lambda^2}\frac{1}{\left(1+{z}/{\lambda}\right)^2},
\end{math}
with the Gouy-Chapman length being $\lambda=\frac{a^2}{2\pi Z
l_{\rm B} N_m}$ as above. While for $a \gg L$ the profiles follow
the Poisson-Boltzmann solution, deviations appear for $a \sim L$
presumably due to the condensation of the counterions inside the
brush. It is interesting to note that all the profiles in Fig.
\ref{fig:density_profile}c collapse on the Poisson-Boltzmann
solution for $z>5\lambda$. Although not attempted here, we note
that more elaborate solutions of the Poisson-Boltzmann equation
suited to the brush geometry can be used to account for the
profiles obtained via simulation \cite{Santan}.

The counterion density profiles appear to develop a depletion in
the vicinity of the surface for $a > 0.55L$, as can be seen in
Figs. \ref{fig:density_profile}a and \ref{fig:density_profile}b.
This feature is absent for $a < 0.55L$, presumably due to the
emergence of the nematic ordering. In fact, one observes that the
depletion is present so long as the $\cos \theta=0$ part of the
histogram for the rods is non-vanishing. This suggests that the
singular behavior of the density profile of the rods near the
surface (described above) might cause the depletion of the
counterions by way of excluded volume interaction: as long as some
of the rods are lying on the surface, they jam the space near the
surface and deplete the counterions. The range of the depletion in
the profiles of Figs. \ref{fig:density_profile}a and
\ref{fig:density_profile}b ($\sim 2 \sigma$) is consistent with
this picture \cite{Asakura}.

In conclusion, we have studied the collective behavior of rod-like
polyelectrolytes that are end-grafted on a substrate, in the
presence of mono- and trivalent counterions. We have demonstrated
that the interplay between condensation of counterions and the
repulsion of the rods can bring about a number of different
regimes, while multivalent counterions can introduce correlations
that lead to a structured collapse of the brush. The emergence of
such patterns in the rod conformations---of staggered type for
monovalent counterions and with ``tepee frameworks'' in case of
the trivalent---seems to originate from similar free energy
minimizing effects, and might have practical applications, in the
design of DNA chips.

\acknowledgements

It is a pleasure to acknowledge fruitful discussions with B.
Bozorgui, M. Grubb, J.-F. Joanny, R.R. Netz, D. Sackett, and J.
Ulstrup.



\begin{thebibliography}{99}

\bibitem{Moreira}
A.G. Moreira and R.R. Netz, in {\it Electrostatic Effects in Soft
Matter and Biophysics}, edited by C. Holm, P. Kekicheff, and R.
Podgornik (Kluwer, Boston, 2001); A.Yu. Grosberg, T.T. Nguyen, and
B.I. Shklovskii, Rev. Mod. Phys. {\bf 74}, 329 (2002).

\bibitem{DNA}
J. Widom and R.L. Baldwin, Biopolymers {\bf 22}, 1595 (1983); V.A.
Bloomfield, Biopolymers {\bf 31}, 1471 (1991).

\bibitem{cyto}
T.D. Pollard and J.A. Cooper, Annu. Rev. Biochem. {\bf 55}, 987
(1986); P.A. Jamney {\em et al.}, Nature (London) {\bf 345}, 89
(1990); G. Isenberg, Semin. Cell Div. Biol. {\bf 7}, 707 (1996).

\bibitem{gerard}
G.C.L. Wong {\em et al.}, Phys. Rev. Lett. {\bf 91}, 018103
(2003).

\bibitem{Gelbart}
K.-C. Lee {\em et al.}, Phys. Rev. Lett. {\bf 93}, 128101 (2004).

\bibitem{Bruinsma}
I. Borukhov, R.F. Bruinsma, Phys. Rev. Lett. {\bf 87}, 158101 (2001).

\bibitem{Fazli}
H. Fazli, R. Golestanian, and M. R. Kolahchi, Phys. Rev. E. {\bf 72}, 011805 (2005).

\bibitem{PEbrush}
For a recent review, see: J. R\"uhe {\em et al.}, Adv. Polym. Sci.
{\bf 165}, 79 (2004).

\bibitem{Sataric}
M. V. Sataric and J. A. Tuszynski,
Phys. Rev. E  {\bf 67}, 011901 (2003).

\bibitem{MT}
Z. Wang and M.P. Sheetz, Biophys J. {\bf 78}, 1955 (2000).

\bibitem{Safinya}
D. J. Needleman {\em et al.}, Phys. Rev. Lett.  {\bf 93}, 198104 (2004).

\bibitem{DNAchips}
A. Halperin, A. Buhot, and E. B. Zhulina, Biophys J. {\bf 89}, 796
(2005).

\bibitem{Mikala}
H. Wackerbarth {\em et al.} Angew. Chem. Int. Ed. {\bf 43}, 198
(2004); Langmuir {\bf 20}, 1647 (2004).

\bibitem{Tirrel}
P. Guenoun {\em et al.}, Phys. Rev. Lett. {\bf 81} 3872 (1998).

\bibitem{Lekner}
J. Lekner, Physica A {\bf 176}, 485 (1991).

\bibitem{Belloni}
M. Roger {\em et al.}, Eur. Phys. J. E {\bf 9}, 313 (2002).

\bibitem{Manning}
G.S. Manning, J. Chem. Phys. {\bf 51}, 954 (1969); F. Oosawa,
Biopolymers {\bf 6}, 134 (1968).

\bibitem{Nematic}
L. Onsager, Ann. N.Y. Acad. Sci. {\bf 51}, 627 (1949).

\bibitem{FII}
B.-Y. Ha and A.J. Liu, Phys. Rev. Lett. {\bf 79}, 1289 (1997); M.
Kardar and R. Golestanian, Rev. Mod. Phys. {\bf 71}, 1233 (1999).

\bibitem{Santan}
For a related study in the context of flexible polyelectrolyte
brushes see: C.D. Santangelo and A.W.C. Lau, Eur. Phys. J. E {\bf
13}, 335 (2004).

\bibitem{Asakura}
S. Asakura and F. Oosawa, J. Chem. Phys. {\bf 22}, 1255 (1954).

\end{thebibliography}
\end{document}